\documentclass[twocolumn]{aa}
\usepackage{graphicx}
\usepackage{epsfig}
\begin{document}
   \title{Radio Recombination Lines from the Starburst Galaxy NGC 3256}

   \titlerunning{Recombination Lines from NGC 3256}

   \author{A.L. Roy\inst{1,2,3,4,5}, 
           W.M. Goss\inst{4}, 
           Niruj R. Mohan\inst{6} \and
           K.R. Anantharamaiah\inst{5}\fnmsep\thanks{deceased}
          }

   \authorrunning{A. L. Roy et al.}

   \institute{$^{1}$Max-Planck-Institut f\"ur Radioastronomie, Auf dem
              H\"ugel 69, 53121 Bonn, Germany\\
              $^{2}$Geod\"atisches Institut der Universit\"at Bonn,
              Nussallee 17, 53115 Bonn, Germany\\
              $^{3}$Australia Telescope National Facility, PO Box 76,
              Epping 1710, NSW, Australia\\
              $^{4}$NRAO, PO Box O, Socorro, NM 87801, USA\\
              $^{5}$Raman Research Institute, CV Raman Ave,
              Sadashivanagar, Bangalore 560080, India\\
              $^{6}$Institut d'Astrophysique de Paris,
              98bis Boulevard Arago, 75014 Paris, France}

   \date{Received ; accepted}

   \abstract{
     
We have detected the radio recombination lines H91$\alpha$ and H92$\alpha$
with rest frequencies of 8.6 GHz and 8.3 GHz from the starburst
nucleus NGC 3256 at an angular resolution of $16.4'' \times 9.6''$
using the Australia Telescope Compact Array and at an angular
resolution of $12.0'' \times 2.9''$ using the VLA.  The line was detected
at $\sim 1$~mJy~beam$^{-1}$ peak with a width of 160~km~s$^{-1}$ with the
ATCA and at $\sim 0.5$~mJy~beam$^{-1}$ peak with a width of 114~km~s$^{-1}$ 
with the VLA.  Modelling the line emitting region
as a collection of H~II regions, we derive constraints on the required
number of H~II regions, their temperature, density, and distribution.  We
find that a collection of 10 to 300 H~II regions with temperatures of
5000 K, densities of 1000~cm$^{-3}$ to 5000~cm$^{-3}$ and
diameters of 15~pc produced good matches to the line and
continuum emmission.  The Lyman continuum production rate required to
maintain the ionization is $2 \times 10^{52}~\mathrm{s}^{-1}$ to
$6 \times 10^{53}~\mathrm{s}^{-1}$, which
requires 600 to 17000 O5 stars to be produced in the starburst.

   }

   \maketitle
%

\section{Introduction}

Starburst activity is one of the more spectacular events of
galaxy evolution.  Star formation is usually slow and well
regulated but occasionally runs away converting much of the
interstellar medium (ISM) in the host galaxy into stars in a
short-lived phase.  Many presently normal galaxies
may once have gone through a starburst phase, which would enrich
and maybe redistribute the ISM.  Understanding starbursts
is therefore important for understanding how galaxies
came to be as they are at present.

Radio recombination lines (RRLs) offer a useful tool for probing
compact nuclear starbursts since radio wavelengths pass unattenuated
through the obscuring dust that hampers optical and near-infrared 
studies of starbursts.
RRL observations also provide dynamical information with unprecedented
resolution, because the lines occur at conveniently high frequencies
where interferometers provide arcsec resolution.

The potential for detecting and exploiting extragalactic RRLs was
shown by Shaver (1978) and shortly thereafter RRLs were detected from
the starbursts in M 82 and NGC 253 in the late 1970's (Shaver et
al. 1977, Seaquist \& Bell 1977).  These
two have since been studied over a wide range of frequencies,
yielding several important constraints on the physical state and
kinematics in the nuclear regions (e.g. Anantharamaiah \& Goss 1997;
Rodriguez-Rico et al. 2004, in prep).

Following those first two detections came a period of surveys that
produced no further detections (Churchwell \& Shaver 1979; Bell \&
Seaquist 1978; Bell et al. 1984).  A renewed effort during the early
1990s using the VLA with improved sensitivity detected RRLs near 8.6~GHz 
from several bright starburst galaxies at a level an order of
magnitude weaker than the first two detections.  These new detections
are NGC 660, NGC 1365, NGC 2146, NGC 3628, NGC 3690, NGC~5253, M 83, IC 694,
Arp 220, Henize~2-10 (Anantharamaiah et al. 1993; Zhao et al. 1996; 
Phookun et al. 1998; Mohan et al. 2001), NGC~1808 (Mohan 2002), and
NGC~4945 at mm wavelengths (Viallefond, private communication).  

During a recent RRL survey using the Australia Telescope Compact Array (ATCA),
we have made three new detections: NGC 3256, NGC 4945 and the Circinus galaxy.
Here, we report the detection of NGC 3256, and the other two 
will be presented in later papers.

NGC 3256 is a pair of colliding disk galaxies that are partly merged and
display spectacular tidal tails and disrupted morphologies (e.g.  Schweizer
1986).  At an inferred distance of 37 Mpc ($v_{\mathrm{r}} =
(2781~\pm~24)$~km~s$^{-1}$ from optical emission lines; de Vaucouleurs et al.
1991), the FIR luminosity is $1.9\times10^{11} L_{\odot}$ following the method
of Helou et al. (1985), making it one of the most luminous galaxies with a
recession velocity less than 3000~km~s$^{-1}$.  The molecular gas mass,
inferred from CO emission, is extremely large ($3\times10^{10} M_{\odot}$;
Sargent et al. 1989).  The far-infrared (FIR) colours (Rowan-Robinson \&
Crawford 1989), and the near-infrared colours (Glass \& Moorwood 1985) are
typical of starburst galaxies. The Br$\gamma$ and [Fe~II] luminosities imply a
high-mass star formation rate of $0.74~M_{\odot}~\mathrm{yr}^{-1}$ or a total
star formation rate of $3.9~M_{\odot}~\mathrm{yr}^{-1}$ and a SN rate of
$0.35~\mathrm{yr}^{-1}$ in the nuclei (Kotilainen et al. 1996), which is
similar to the total star formation rate of 3~$M_{\odot}~\mathrm{yr}^{-1}$ for
the Milky Way (Telesco 1988).  The radio continuum is predominantly
non-thermal betwen 408 MHz and 5000 MHz with a spectral index of -0.77
(PKSCAT90).  The system displays the normal radio-FIR correlation which is
characteristic of normal and star-forming galaxies.  VLA observations at 6 cm
with 4" resolution by Smith \& Kassim (1993) show emission over 30" and arms
of diffuse emission extending out towards the giant tidal arms seen in H~I
(English et al. 2003).  At higher resolution (2"), Norris \& Forbes (1995)
resolve the nucleus into two equal components, which they argue are the two
nuclei of the progenitor galaxies, both undergoing starbursts.  From the
synchrotron luminosity of the nuclei, they derive a supernova rate of
0.3~yr$^{-1}$ in each nucleus, consistent with estimates from Br$\gamma$ and
[Fe~II].  X-ray emission from both nuclei (Lira et al. 2002) is consistent
with that from low-luminosity active galactic nuclei (Neff, Ulvestad \&
Campion 2003).  In summary, the system is a pair of gas-rich disk galaxies
that are colliding and hosting a spectacular burst of star formation and
AGN activity.

We adopt $H_{0} = 75~\mathrm{km~s^{-1} Mpc^{-1}}$, $q_{0} = 0.5$ and 
$\Lambda = 0$, and give velocities in the heliocentric frame using
the optical velocity definition throughout this paper.


\section{Observations}

{\bf Australia Telescope Compact Array (ATCA)} (Frater \& Brooks 1992) 
observations were made  during 1994 Oct
22 to 24 and 1995 Aug 05.  The ATCA was configured with five antennas on
an E-W track with baselines between 46 m and 750 m.

We observed simultaneously the lines H91$\alpha$ and H92$\alpha$ with
rest frequencies of 8584.82~MHz and 8309.38~MHz, and recorded two
orthogonal linear polarizations.  We used 64 spectral channels across
a 64 MHz bandwidth covering each transition, which corresponds to a
velocity coverage of $2270~\mathrm{km~s^{-1}}$ with 
$35~\mathrm{km~s^{-1}~channel^{-1}}$.  We integrated for 24.2~h on-source.

Calibration and imaging were done using the AIPS software, using standard
methods.  The flux-density scale assumed that PKS B1934-638 has a flux density
of 2.81~Jy at 8236~MHz and 2.66~Jy at 8474~MHz, based on the Baars et al.
(1977) flux-density scale.  A phase calibrator was observed every half hour to
correct the instrumental phase response.  A bandpass calibrator was observed
every few hours for correcting the instrumental frequency response (bandpass).
Phase corrections obtained from self calibration of the continuum source were
applied to the spectral line data.  Continuum emission was subtracted from the
line data using the method UVLSF (Cornwell, Uson \& Haddad 1992), in which the
continuum is determined for each baseline by a linear fit to the spectrum.
The final continuum and line images were made using robust weighting
(robustness 1) of the $(u, v)$ data to achieve near-maximum possible
signal-to-noise ratio with resolution of $16.4'' \times 9.6''$ in PA
$-14^{\circ}$, which was 20~\% higher than with natural weighting.  The rms
noise level in the spectrum was $130~\mathrm{\mu Jy~beam^{-1}}$ per 1.0~MHz
channel after averaging together the two transitions and two polarizations.

Uncertainties on the flux densities have an 11~\% rms random multiplicative
component due to flux-density bootstrapping and atmospheric opacity, a
0.13~mJy rms random additive component due to thermal noise in a 1~MHz channel
or 0.16~mJy rms in the continuum image, and a systematic multiplicative
component of 11~\% rms due mainly to the uncertainty in the Baars et al.
flux-density scale.

{\bf Very Large Array (VLA)} observations were made in the CnB configuration
on 1998 Nov 05 and 06 for 8.4~h with
a beamsize of $3.9'' \times 2.1''$ in PA $5^{\circ}$ and in C-array on 
1998 Dec 07 and 1999 Jan 15 for 3.8 h total, with a beamsize
of $12.0'' \times 2.9''$ in PA 10$^{\circ}$ (natural weight), 
to confirm the ATCA observation.

We observed H92$\alpha$ and recorded two orthogonal circular
polarizations.  We used 15 spectral channels across a 25~MHz bandwidth
in 1998 Nov, which corresponds to a velocity coverage of
$910~\mathrm{km~s^{-1}}$ with $61~\mathrm{km~s^{-1}~channel^{-1}}$ 
We used 15 spectral channels across a 50~MHz bandwidth in 1998 Dec 
and 1999 Jan, providing twice the velocity coverage.

Data reduction followed the same procedure as for the ATCA data 
except as follows.  The flux-density scale assumed that 3C 286 had a flux
density of 5.23~Jy at 8235~MHz (Baars et al. 1977),
the phase calibrator served also as bandpass calibrator, and we used
natural weighting for the line detection and uniform
weighting for best resolution in the continuum image.  The rms noise level in
the 1998 Nov spectrum was $81~\mathrm{\mu Jy~beam^{-1}}$ per 1.562 MHz
channel, and for the combined 1998~Dec and 1999~Jan spectrum was
$140~\mathrm{\mu Jy~beam^{-1}}$ per 3.125 MHz channel after averaging together
the two polarizations.  Since the sensitivity of the 1998 Nov observation was
better than the later two observations, and the data could not easily be
combined due to their different spectral resolution; the results presented
here are based on the 1998 Dec + 1999 Jan data since the 1998 Nov
data did not detect line emission.

The observational parameters are summarized in Table 1.

{\bf The Continuum Spectrum} was required for separating the thermal and
non-thermal continuum components during modelling, and so we used archival 
VLA B-array continuum observations
at 1.4~GHz and 4.8~GHz to complement the ATCA 8.4~GHz measurement.
The VLA observations were made on 1990 Oct 05 in project AS412.  The
flux-density scale was calibrated assuming 3C~48 had a flux density of
15.62~Jy at 1.4~GHz and 5.49~Jy at 4.8~GHz.  We tapered the 4.8 GHz array to
match the beamsize of the the ATCA at 8.4 GHz ($19.9'' \times 10.8''$).  Since
the full-resolution array at 1.4~GHz provided a larger beamwidth ($27.6''
\times 11.3''$), we estimated a correction factor to account for the beam
mismatch by tapering the 4.8~GHz data to the beamsize of the 1.4~GHz image and
calculating the ratio of the peak flux densities measured from the 4.8~GHz
images at the two resolutions.  This factor, 1.16, was used to scale down the
measured 1.4~GHz peak flux density to give an expected value for the smaller
beamsize.  Since the line emission was unresolved in the ATCA observation, we
used the brightness in the continuum images at the position of the peak of
the line emission for the continuum flux density for estimating the continuum
spectral index.  The results are given in Table 2.

\section{Results}

{\bf The ATCA continuum and line images,} integrated spectrum, and
position-velocity diagrams are shown in Figs. 1 to 3.  The measured line
and continuum parameters are given in Table 2. 

The continuum image shows a single, slightly resolved component.  The double
nuclei detected by Norris \& Forbes (1995) are 4.9'' apart and so
were not separated by the relatively large ATCA beam.  The continuum emission
is predominantly non-thermal, with a spectral index of -0.88 between 1.5~GHz
and 8.3~GHz (Table 2).

Line emission was detected in the nuclear region, nearly coincident with the
peak continuum emission.  The line emission region was only slightly resolved
with the $16.4'' \times 9.6''$ beam, which implies a size of $\leq 2.9$~kpc.
The offset between the line and continuum peak in Fig 1 is perhaps 
significant as it is seen also in the VLA observations.
The low-level line emission to the north outside the 
continuum contours is probably not significant and was also not confirmed
by the VLA observations.

The H91$\alpha$ + H92$\alpha$ spectrum integrated over the compact
line-emitting region near the continuum maximum shows a clear line detection
at $7.3~\sigma$ with centroid at 2772~km~s$^{-1}$, compared to the systemic 
velocity of 2781~km~s$^{-1}$.  The line FWHM is 160~km~s$^{-1}$.  

The position-velocity diagram (Fig 3) shows the strongest emission near the
continuum peak and close to systemic velocity with some complex dynamics in
the declination cuts where they intersect the SE extension.  The figure shows
emission with marginal significance ($3.3~\sigma$) detected in a southern
extension at dec = $-43^{\circ}~54'~28''$.

{\bf The VLA continuum image and line spectrum} are shown in Figs. 4 and 5,
and the measured line and continuum parameters are given in Table 2.

The VLA continuum image with line emission overlayed from 98Dec + 99Jan
is shown in Fig 4.  It shows a $4.0~\sigma$ peak of line emission 
within the region of continuum emission.

The VLA continuum image at high resolution on 98Nov in Fig 5 
has four times higher resolution than the
ATCA image and resolves the continuum emission into two nuclei of nearly equal
strength separated by 4.8'' (870 pc), consistent with the ATCA continuum image
by Norris \& Forbes (1995).  The nuclei are resolved, with FWHM diameters of
1.8'' (320 pc) and 2.5'' (450 pc).  These also are detected in 
H$_{2}$, Br$\gamma$, and [Fe II] (Moorwood \& Oliva 1994).  Surrounding
the two nuclei is a broad low-brightness `S'-shaped region of radio emission
that may be the base of the tidal arms (Kotilainen et al. 1996).

The H92$\alpha$ line emission (Fig 6) was detected at 
0.56~mJy~beam$^{-1}$ ($4.0~\sigma$) in the 98Dec + 99Jan observation,
which confirms the ATCA detection.  The peak line strength detected by
the VLA appears to be less than that detected by the ATCA.  Tapering the VLA
data to match the larger beamsize of the ATCA yielded a peak line strength
of 0.49~mJy~beam$^{-1}$ and so did not reduce the apparent inconsistency.
However, the difference between the VLA and ATCA peak line strengths is only
$0.9~\sigma$ and so the measurements are formally consistent with each
other.  The line was not detected in the 98Nov observation at either full
resolution with natural weight or tapered to match the ATCA beam and we can
offer no explanation for this.  The $3~\sigma$ upper limit of
0.24~mJy~beam$^{-1}$ is lower than the line detections made with the ATCA
and with the VLA in the 98Dec + 99Jan observation, but the differences have
formal significance of only $1.0~\sigma$ and $2.5~\sigma$.  The three
observations are therefore formally consistent with each other.

Emission is centred near 2800~km~s$^{-1}$, near the systemic velocity.  This
narrow RRL component is coincident in velocity with the H~I absorption towards
the nucleus seen by English et al. (2003), which spans 2700~km~s$^{-1}$ to
2960~km~s$^{-1}$.

\begin{figure}
\centering
\includegraphics[width=8cm]{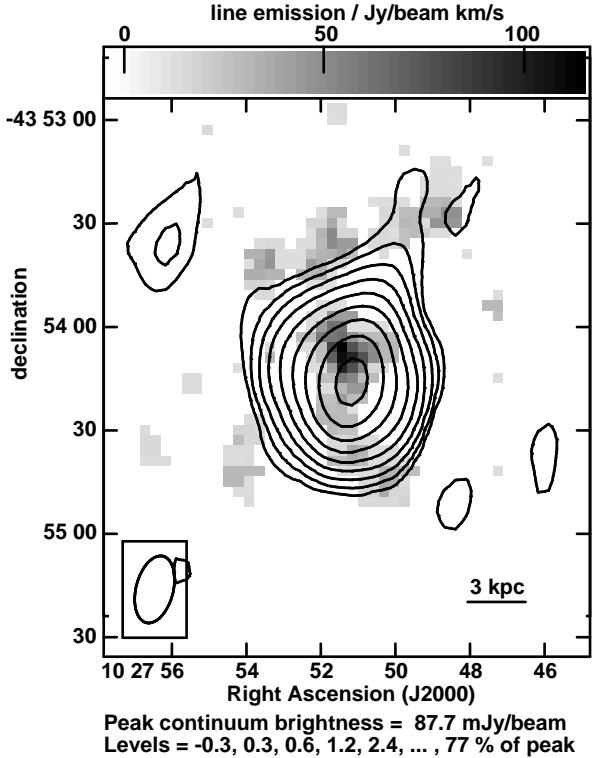}
\caption{
ATCA 8.4 GHz continuum image of NGC 3256 observed 1994 Oct + 1995 Aug
(contours), superimposed on the grey scale moment 0 image showing 
H91$\alpha$ + H92$\alpha$ line emission.  Beamsize is 
$16.4'' \times 9.6''$, rms noise is 0.14\,mJy\,beam$^{-1}$.
}
\label{ATCACont}
\end{figure}

\begin{figure}
\centering
\includegraphics[width=8cm]{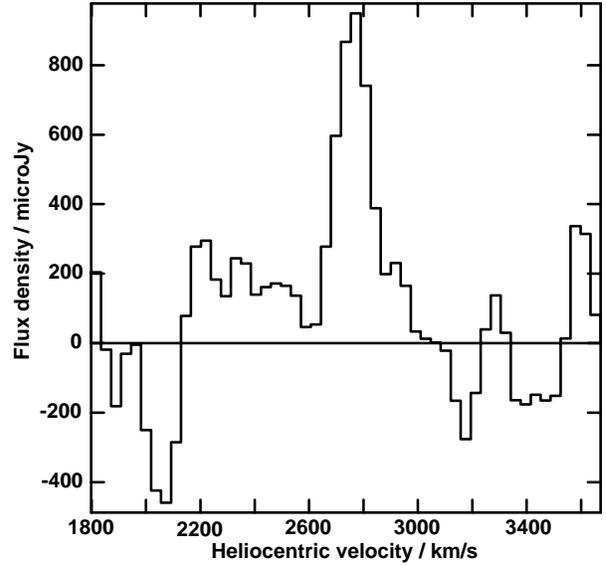}
\caption{
ATCA H91$\alpha$ + H92$\alpha$ line profile integrated over the
line-emitting region in NGC 3256, observed 1994 Oct + 1995 Aug.
Region of integration is a box of size $24'' \times 27''$ centred
on RA 10 27 51.303  dec -43 54 05.7.
}
\label{ATCAISPEC}
\end{figure}

\begin{figure}
\centering
\includegraphics[width=8cm]{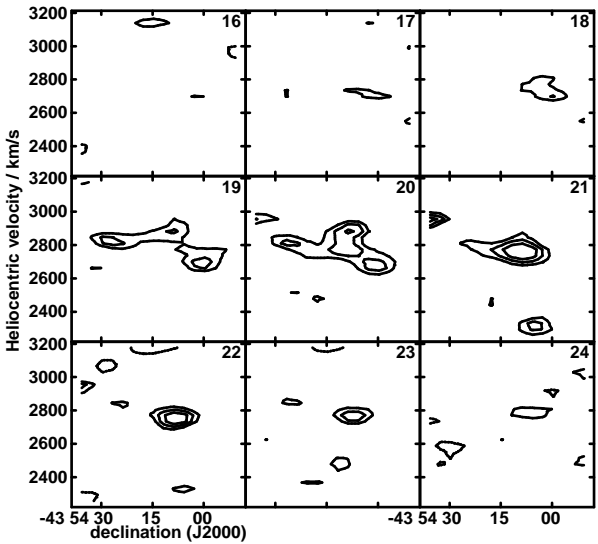}
\caption{
ATCA position-velocity diagram in NGC 3256 along declination cuts at
RA = 10 27 50.33 to 10 27 52.55 (J2000)
observed 1994 Oct + 1995 Aug.  Contour levels are -0.5, -0.4, -0.3,
0.3, 0.4, 0.5 mJy beam$^{-1}$.  The peak flux density is 0.66~mJy~beam$^{-1}$
and the rms noise is 0.13~mJy~beam$^{-1}$.
}
\label{ATCAPV}
\end{figure}

\begin{figure}
\centering
\includegraphics[width=8cm]{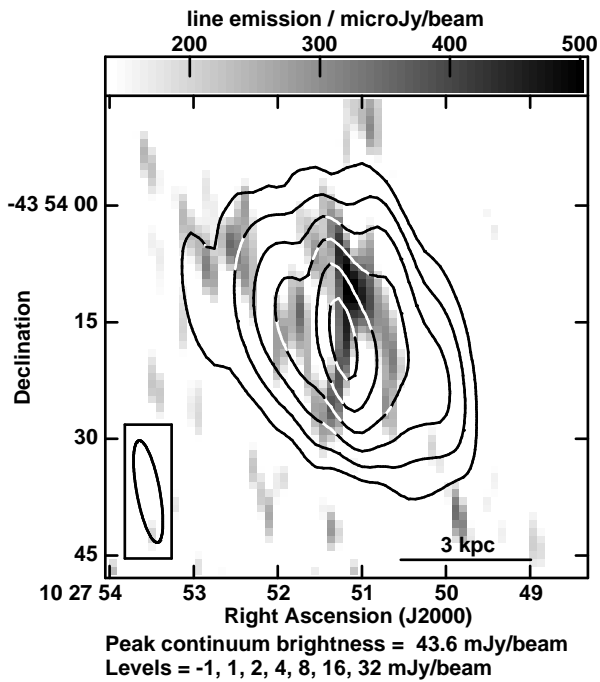}
\caption{
VLA 8.4 GHz continuum image with natural weight of NGC 3256 observed  
1998 Dec + 1999 Jan (contours), superimposed on the grey scale image of 
the channel at 2817~km~s$^{-1}$ showing H92$\alpha$ line emission.
Beamsize is $12.0'' \times 2.9''$, rms noise is 0.14\,mJy\,beam$^{-1}$.  The
grey scale is truncated to white below 1~$\sigma$.
}
\label{VLAContLine}
\end{figure}

\begin{figure}
\centering
\includegraphics[width=8cm]{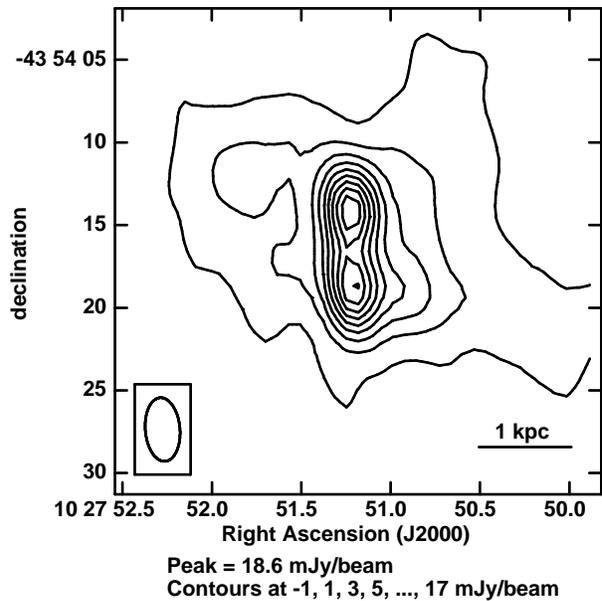}
\caption{
VLA 8.4 GHz continuum image at higher resolution, 
with uniform weight of NGC 3256 observed 1998 Nov, showing the double nucleus.
Beamsize is $3.9'' \times 2.1''$, rms noise is 0.10\,mJy\,beam$^{-1}$.
}
\label{VLACont}
\end{figure}

\begin{figure}
\centering
\includegraphics[width=8cm]{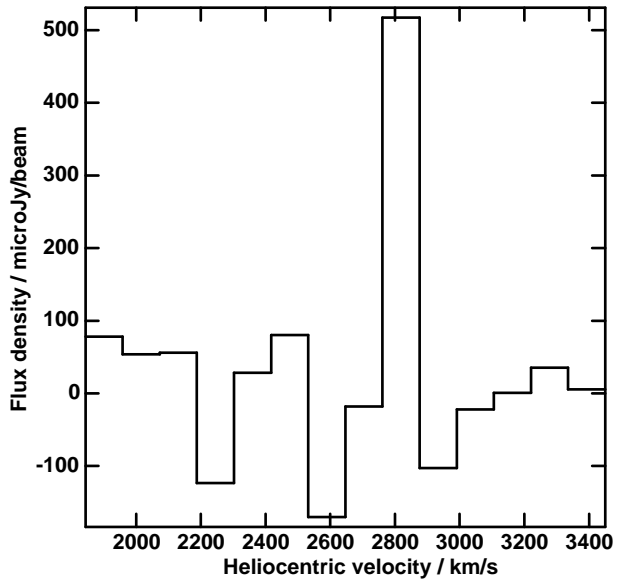}
\caption{
VLA H92$\alpha$ line profile at the peak of the line emission
in the nucleus NGC 3256, observed 1998 Dec + 1999 Jan, observed with a 50~MHz
bandwidth.  Beamwidth is 
12.0''~$\times$~2.9'' in P.A. $4^{\circ}$ and does not resolve the
double nuclei.
}
\label{VLASPECBROAD}
\end{figure}

\begin{table*}
\caption[]{Observational Parameters for NGC 3256.}
\label{ObsLine}
\begin{center}
\begin{tabular}{llll}
\hline
\hline
\noalign{\smallskip}
Telescope                   &  ATCA                        & VLA  & VLA \\
\noalign{\smallskip}
\hline
\noalign{\smallskip}
Date of observing           &  1994 oct 24 and 1995 aug 05 & 1998 nov 05/06 &
1998 dec 07 + 1999 jan 15  \\
Array configuration         &  750C (1994 oct) 375 (1995 aug) & CnB  & C \\
No. antennas                &  6                           & 27 & 27 \\
Transitions                 &  H92$\alpha$ \& H91$\alpha$  & H92$\alpha$ & H92$\alpha$ \\
Freq. at band centre        &  8236~MHz and 8474~MHz       & 8233.55~MHz &
8235.1~MHz \\
Beam size &  $16.4'' \times 9.6'' $ at -$14^{\circ}$
         &  $3.9'' \times 2.1''$ in PA 5$^{\circ}$ $^a$  &
         $12.0'' \times 2.9''$ in PA 10$^{\circ}$ \\
         & & $5.7'' \times 2.6''$ in PA 1$^{\circ}$ $^b$ \\

Spectral channels           &  64 & 15 & 15 \\
Total bandwidth             &  64~MHz = 2270~km~s$^{-1}$   
                            &  25~MHz = 910~km~s$^{-1}$ 
                            &  50~MHz = 1820~km~s$^{-1}$ \\
Spectral resolution         &  1~MHz = 35~km~s$^{-1}$ 
                            &  1.562~MHz = 56~km~s$^{-1}$ 
                            &  3.125~MHz = 112~km~s$^{-1}$ \\
$V_{\mathrm{helio,optical}}$&  2775~km~s$^{-1}$            & 2759~km~s$^{-1}$
& 2683~km~s$^{-1}$ \\
Integration time            &  24.2~h                      & 8.4~h & 3.8 h \\
$T_{\mathrm{sys}}$          &  44 K                        &  35 K (nominal) &
35 K (nominal) \\
Polarization                &  dual linear                 & dual circular &
dual circular  \\
Phase calibrator            &  1104-445               & 1104-445 & 1107-448 \\
Bandpass calibrators        &  1104-445, 1921-293, 2251+158& 1104-445 &
1104-445 \\
Flux density cal            &  1934-638                    & 3C 286 &
3C 286 \\
Assumed flux density        &  2.81 Jy at 8236 MHz & 5.23 Jy at 8235 MHz &
5.23 Jy at 8235 MHz \\
                            &  2.66 Jy at 8474 MHz \\
Noise per image channel     &  0.13~mJy~beam$^{-1}$          &
0.081~mJy~beam$^{-1}$  & 0.14~mJy~beam$^{-1}$ \\
\noalign{\smallskip}
\hline
\end{tabular}
\end{center}
$^{a}$ Continuum image made with uniform weight \\
$^{b}$ Line image made with natural weight \\
\end{table*}

\begin{table*}
\caption[]{Observational Results for NGC 3256.}
\label{ResultLine}
\begin{center}
\begin{tabular}{lllll}
\hline
\hline
\noalign{\smallskip}
Parameter     &  ATCA $^a$  & VLA (north nucleus) & VLA (south nucleus) & VLA
(98Dec+99Jan) \\
\noalign{\smallskip}
\hline
\noalign{\smallskip}
\bf{Continuum properties} \\
Peak position RA (J2000)     &  10~27~51.182  & 10 27 51.2199  & 
10 27 51.1946  & 10 27 51.2296 \\
Peak position declination    &  -43~54~15.89  & -43 54 13.79 
& -43 54 18.61 & -43 54 17.26 \\
Peak flux density (mJy~beam$^{-1}$) &  $89.1 \pm 13$ & $15.9 \pm 2.5$ 
& $15.1 \pm 2.4$ & $43.4 \pm 6.8$ \\
Integrated flux density (mJy) &  $152 \pm 24$ & $25.4 \pm 4.0$ &
$27.8 \pm 4.3$ & $156 \pm 24$ \\
Deconvolved size            &  $13.3'' \times 10.6''$ at $47^{\circ}$ & 
$2.3'' \times 1.4''$ at $8^{\circ}$ &
$2.8'' \times 2.2''$ at $42^{\circ}$ &
$5.0'' \times 3.8''$ at $10^{\circ}$ \\
\\
\bf{Line properties} \\
Peak line flux density (mJy~beam$^{-1}$)     &  $0.95 \pm 0.27$  & $< 0.24$
 & $< 0.24$ & 
$0.56 \pm 0.32$ \\
Velocity dispersion (km~s$^{-1}$)           &  $87 \pm 26$  & --      & -- &
$48 \pm 25$ \\
Integrated line flux (Jy~km~s$^{-1}$)      &  $0.16 \pm 0.05$ & --
  & -- &
$0.05 \pm 0.03$ \\
Integrated line flux ($10^{-23}$~W~m$^{-2}$) & $4.3 \pm 1.3$
& --
& -- &  $1.4 \pm 0.8$ \\
Cont. flux density over line region (mJy)  & $116 \pm 18$ & -- & -- & $ 19 \pm
3 $ \\
Centroid helio. optical vel. (~km~s$^{-1}$)  &  $2772 \pm 22$   &  --
& -- & $2817 \pm 36$ \\

No. of beam areas where line is observed & 0.1 & -- & -- & 1.0 \\

Line width (FWHM) in km~s$^{-1}$)     &  $161 \pm 48$   &  --        & --   
         &  $114 \pm 60$ \\
Distance & 37~Mpc & & \\
\\
\bf{Other Measurements}\\
Continuum flux density at: \\
1.49~GHz (mJy~beam$^{-1}$) &  211$^b$ \\
4.90~GHz (mJy~beam$^{-1}$) &  119     \\
8.30~GHz (mJy~beam$^{-1}$) &  49      \\
Estimated thermal em. at 8.3 GHz (mJy) & 26 & \\
Br$\gamma$ flux (W~m$^{-2}$)   &  $5.3 \times 10^{-17}$ $^c$      \\

\noalign{\smallskip}
\hline
\end{tabular}
\end{center}
$^{a}$ These ATCA measurements were used in the radiative transfer modelling\\
$^{b}$ Beamwidth = $19.9'' \times 10.8''$ \\
$^{c}$ 6'' aperture, Kotilainen et al. 1996
\end{table*}


\section{Modelling the Ionized Gas}

Conditions in the ionized gas are constrained by the observed line and
continuum emission.  Following Anantharamaiah et al. (1993), we
consider two simple models: 1) a uniform slab of ionized gas in front
of the central non-thermal continuum source, and 2) a collection of 
H~II regions within the central few-hundred parsecs.

1. The slab model consists of a slab of ionized gas with the same lateral
   size as the line-emitting region, and is characterized by an electron
   temperature $T_{\mathrm{e}}$, electron density $n_{\mathrm{e}}$
   and thickness along the line of sight $l$.  The peak line flux
   density is then given by Eq. 1 of Anantharamaiah et al. (1993) and
   contains contributions from spontaneous emission from the slab,
   stimulated emission amplifying thermal emission within the slab,
   and stimulated emission amplifying the background continuum.
   Thermal continuum emission from the slab is readily derived given
   $T_{\mathrm{e}}$, $n_{\mathrm{e}}$ and $l$.  We found that there
   are no uniform-slab models that fit the observed line and continuum
   flux densities simultaneously and so we do not consider this model further.

2. The collection of H~II regions model consists of 
   a collection of spherical H~II regions, all with the same 
   $T_{\mathrm{e}}$, $n_{\mathrm{e}}$ and linear diameter $l$ embedded in
   a volume of uniform synchrotron emission.  The
   total number of clouds, $N$, is determined by calculating the line
   flux density produced by a single cloud with the given conditions
   and dividing that into the total observed line strength.  Some
   combinations of conditions can be ruled out using the following
   requirements. 1) The volume filling factor of H~II regions,
   calculated by dividing the volume of the line-emitting region by
   $N$ times the volume of an individual H~II region, should not
   exceed unity. 2) Since the line width of a single H~II region is much less
   than the observed line width, a minimum number of H~II regions with
   different velocities must exist inside every beam are within the
   line-emitting region. 3) The peak line flux density of a single
   H~II region should not exceed the observed line flux density.
   4) The total thermal continuum emission from the H~II regions,
   $S_{\mathrm{th}}$, should not exceed that inferred from the Br$\gamma$
   flux.

We considered a grid of models with $T_{\mathrm{e}}$, $n_{\mathrm{e}}$
and $l$ in the ranges 1000~K to 12500~K, 10~cm$^{-3}$ to $10^{6}$~cm$^{-3}$
and 0.01 pc to 100 pc.  Models with 10 to 300 H~II regions, all with
$T_{\mathrm{e}} \sim 5000$~K, $n_{\mathrm{e}} \sim 
10^{3}$~cm$^{-3}$ to $10^{4}$~cm$^{-3}$ and size $l \sim 15$~pc
produced good matches to the line and continuum emission.

Parameters derived for typical allowed models are given in Table 3.

\begin{table*}
\caption[]{Properties of the ionized gas in both nuclei in NGC~3256 derived
  from the ATCA line detection using the Collection of H II Regions model.}
\label{ModelRes}
\begin{center}
\begin{tabular}{ll}
\hline
\hline
\noalign{\smallskip}
Parameter      &  Value \\
\noalign{\smallskip}
\hline
\noalign{\smallskip}
Source                      &  NGC 3256            \\
Electron temperature        &  5000 K              \\
Electron density            &  1000~cm$^{-3}$ to 5000~cm$^{-3}$        \\
Number of HII regions       &  10 to 300                  \\
Size                        &  15~pc                \\
Total ionized gas mass      &  $4\times10^{4}~M_{\odot}$ to
                               $2\times10^{5}~M_{\odot}$ \\
$N_{\mathrm{LyC}}$          &  $2\times10^{52}$~s$^{-1}$ to
                               $6\times10^{53}$~s$^{-1}$    \\
No. O5 stars                &  600 to 17000                 \\
Fraction of thermal continuum at 5 GHz &  1~\% to 7~\%    \\
Filling factor              &  $3\times10^{-4}$ to $8\times10^{-3}$    \\
\noalign{\smallskip}
\hline
\end{tabular}
\end{center}
\end{table*}

\section{Discussion}

The strength of the RRL emission infers a mass of ionized gas of
$4\times10^{4}~M_{\odot}$ to $2\times10^{5}~M_{\odot}$, depending on the model
conditions, which requires a Lyman continuum flux of $2\times10^{52}$~s$^{-1}$
to $6\times10^{53}$~s$^{-1}$ to maintain the ionization.  This flux is
equivalent to the Lyman continuum output of 600 to 17000 stars of type O5.

This ionized gas, if spread over two nuclei of 320~pc and 450~pc diameter has
a surface density of (0.2 to 0.8)~$M_{\odot}$~pc$^{-2}$,
compared to the star formation threshold found by Kennicutt (1989) of (3 to
10)~$M_{\odot}$~pc$^{-2}$ for the total gas mass. 

The molecular gas mass inferred from CO emission by Sargent et al. (1989), is
$10^5$ times larger than the ionized gas mass we derive from the RRLs, and if
the molecular gas is spread over the central kiloparsec, the surface density
of 40000~$M_{\odot}$~pc$^{-2}$ far exceeds Kennicutt's star formation
threshold.  Thus, of the vast quantity of molecular gas present, only a tiny
fraction has been ionized in the starburst.

The expected X-ray luminosity of 17000 stars of type O5, adopting
$10^{26.3}$~W per O5 star (Chlebowski, Harnden, \& Sciortino 1989) is
$3.4 \times 10^{30}$~W, which is a factor 1000 less than the observed
X-ray luminosity from the two compact X-ray components that are 
coincident with the radio components, namely ULX 7(N) and ULX 8(S).
Thus, we confirm the conclusion of Neff, Ulvestad \& Campion (2003)
that the star formation activity is insufficient to account for the
nuclear X-ray emission.

The observed dereddened Br$\gamma$ flux is $1.54\times10^{-16}$~W~m$^{-2}$
(Prestwich et al. 1994) which agrees with that predicted from the measured RRL
strength, (0.3~to~1.5)$\times10^{-16}$~W~m$^{-2}$, depending on the model
parameters.  The dereddened Br$\alpha$ flux is $13.5\times10^{-16}$~W~m$^{-2}$
(Rigopoulou et al. 1996), yielding a ratio of Br$\alpha$ to Br$\gamma$ of 8.8,
which is different from the expected 2.96 assuming case B recombination.  This
discrepancy could be due to the use of different aperture sizes and extinction
models for measuring and dereddening the different transitions, or due to
the simple foreground screen model not being accurate in the case of 
clumpy heavy extinction.


We infer an electron density of 5000~cm$^{-3}$ within the H~II regions
in the central starburst, though values in the range 1000~cm$^{-3}$ to
10000~cm$^{-3}$ are also allowed by the available constraints from the
RRL detection.  This high density at first appears to be in conflict
with the electron density of 300~cm$^{-3}$ derived by Rigopoulou et
al. (1996) derived from the [S~III] line ratios, and with the limit of 
$< 1000$~cm$^{-3}$ established by Carral et al. (1994) using the [O~III]
line ratios.  However, Anantharamaiah et al. (2000) point out that
lower values from the [S III] method may be a selection effect caused by
the insensitivity of the [S III] ratios to densities outside the range
($10^2$ to $10^{3.5}$) cm$^{-3}$ (Houck et al. 1984).

\section{Conclusions}

We have detected H91$\alpha$ and H92$\alpha$ lines in emission in NGC 3256
using the ATCA and the VLA with a width of 160~km~s$^{-1}$ (measured with the
ATCA) and 114~km~s$^{-1}$ (measured with the VLA).  The line emission region
is coincident with the continuum emission originating in the nuclei and its
integrated flux infers a mass of ionized gas of $4\times10^{4}~M_{\odot}$ to
$2\times10^{5}~M_{\odot}$, requiring 600 to 17000 O5 stars to maintain the
ionization.

\begin{acknowledgements}

The National Radio Astronomy Observatory is a facility of the National
Science Foundation operated under cooperative agreement by Associated
Universities, Inc.  The Australia Telescope Compact Array is part of the
Australia Telescope, which is funded by the Commonwealth of Australia for
operation as a National Facility managed by CSIRO.

\end{acknowledgements}

\end{document}